\newcommand{\be}[0]{\begin{equation}}
\newcommand{\ee}[0]{\end{equation}}
\newcommand{\bea}[0]{\begin{eqnarray}}
\newcommand{\eea}[0]{\end{eqnarray}}
\newcommand{\qq}[0]{$\overline q q$~}

\documentclass[12pt]{article}
\usepackage{amssymb,cite,epsfig,graphics,graphicx}
\begin{document}
\begin{titlepage}
\vskip 0.5cm
\begin{flushright}
DCPT-02/38 \\
IPPP-02/19 \\
\end{flushright}
\vskip 1.5cm
\begin{center}
{\Large{\bf Dynamical Generation of Scalar Mesons
}}
\end{center}
\vskip 1cm
\large
\centerline{M. Boglione}
\vskip 0.5cm 
\centerline{and}
\vskip 0.5cm 
\centerline{M.R. Pennington}
\vskip 1cm
\normalsize
\centerline{\it
Institute for Particle Physics Phenomenology,}
\centerline{\it
University of Durham, Durham DH1 3LE, UK\,}
\vskip 1cm
\begin{abstract}
\noindent
Starting with just one bare seed for each member of a scalar nonet,
we investigate when it is possible to generate more than one hadronic 
state for each set of  quantum numbers. 
In the framework of a simple model, we find that in the $I=1$ sector it is 
possible to generate two physical states with the right features 
to be identified with the $a_0(980)$ and the $a_0(1450)$. 
In the $I=1/2$ sector, we can generate a number of 
physical states with masses higher than $1$ GeV, including one with the 
right features to be associated with the $K_0^*(1470)$, but none which can 
be identified with the light $\kappa$ scalar meson.
The $I=0$ sector is the most complicated and elusive: since all outcomes are 
very strongly model dependent, we cannot draw any 
robust conclusion. Nevertheless, we find that in that case too, depending 
on the coupling scheme adopted, the occurrence of numerous states {\it can} 
be achieved. 
This shows that dynamical generation of physical states is a possible solution 
to the problem of accounting for more scalar mesons than can fit in a single 
nonet, as experiments clearly deliver. 
\end{abstract}
{\small \mbox{}\hspace{0.9cm} PACS numbers: 
12.40.Yx, 13.75.Lb, 14.40.Cs, 14.40.Ev}
\end{titlepage}
\setcounter{footnote}{0}
\def\thefootnote{\fnsymbol{footnote}}
\baselineskip = 6mm
\parskip=2mm

%
\section{Introduction}

\vspace{0.3cm}

In the naive quark model picture with three
flavours, quarks and antiquarks are assumed to be bound into states, the
quantum numbers of which are determined by the spin $S$ and the relative
orbital angular momentum $L$ of the \qq system. This leads to the multiplet
structures that can be elegantly described by the $SU(3)$ group of flavor 
symmetry. The masses of hadrons are then related to the constituent masses
of the quarks and simple relations among them are found. 
For instance, the non-strange $\rho$ and $\omega$ vector mesons, both made 
out of up and down quarks, have 
roughly the same mass, whereas the $\phi$, being a pure $s \overline
s$ state has a mass approximately 300 MeV heavier. Furthermore, the mass of
a meson like the $\rho$, made of two constituent quarks, is about 2/3 of
the mass of a proton or a neutron, made of three such quarks.  
However, the simple and successful picture that the quark model delivers does 
not apply to the scalar meson sector: apparently scalars are different.
 First of all 
there are far more scalar mesons  than can be  
accomodated in one conventional nonet, moreover their masses 
turn out to be hundreds of MeV lighter than one would simply deduce from the 
constituent structure of the mesons. 

\noindent
In Ref.~\cite{torn}, Tornqvist presented a model in which
the central focus is to consider the loop 
contributions given by the hadronic intermediate 
states that each meson can access: it is via these hadronic loops that the
bare states become \lq\lq dressed\rq\rq~ and, in the case of scalar mesons,  
hadronic loop contributions totally dominate the dynamics of
the process.  
He shows that the mass shift, which 
is a direct consequence of the presence of strongly coupled
hadronic intermediate states, is so dramatic that it completely spoils the 
one--to--one correspondence between the resonances we observe and the 
underlying 
constituent structure. Though we follow Tornqvist's modelling quite closely,
very similar models have been considered by 
van Beveren {et al.}~\cite{vanbev}, Geiger and 
Isgur~\cite{geiger-isgur} and by Oller and Oset \cite{oller-oset} among others.

\noindent
In this paper, following and extending the method of Tornqvist and Roos 
\cite{roos}, we will investigate the possibility of generating, in 
the scalar sector, more 
than one 
state with the same quantum numbers, by initially inserting only one 
\lq\lq bare seed\rq\rq.
We will show that the outcome depends on the kinematics of the  intermediate 
channels: crucially, on the number and position of each threshold 
opening and on the strength of their individual couplings. Therefore, 
every case has to be considered separately and it is not possible to reach 
one common conclusion for all the members of the scalar meson family.

\newpage

\section {The model of hadronic dressing}

\vspace{0.3cm} 

\noindent
We start by considering a simple model in which all bare meson states
belong to ideally mixed quark multiplets. 
We call $n \overline n$ the 
nonstrange light state and suppose that substituting a strange quark for a 
light one increases the mass of the state by $\Delta m_s \simeq 150$ MeV.

\begin{figure}[b]
\begin{center}
\mbox{~\epsfig{file=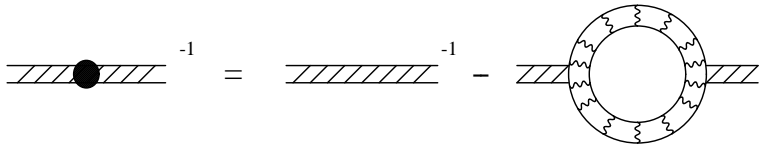,angle=0,width=10cm}}
\caption{\label{prop} \footnotesize{Pictorial representation of the full  
propagator in Eq.~(\ref{fullprop}).
}}
\end{center}
\end{figure}

\noindent
The bare propagator for each of these bound states will be of the form 
\be
P \,=\, \frac{1}{{\cal{M}} _0^2 - s } \; ,
\ee
with a pole on the real axis, corresponding to a non decaying
state; for example for the vector $I=0$ state
\[ |\phi \rangle _0 = |s \overline s \rangle . \]
If we now assume that the experimentally observed hadrons are obtained from 
the bare states ($n \overline n$, $s \overline n$, $s \overline s$, ...)
by dressing them with hadronic interactions, the propagator becomes
\be 
P(s) \,=\, \frac{1}{{\cal{M}} ^2(s) - s -i \, {\cal{M}} (s) \, 
\Gamma (s)}  \; ,
\label{fullprop}
\ee
where a sum over all hadronic interactions (in the loop) is implicit 
(see Fig. \ref{prop}). The pole then moves in the complex $s$-plane.
%
%
The corresponding vector state can be decomposed as
\be 
|\phi \rangle = \sqrt{1-\epsilon ^2}|s \overline s \rangle + \epsilon _1 
|K \overline K \rangle 
 + \epsilon _2 |\rho \pi \rangle  + ...
\ee
where calculation would give \( \epsilon ^2 = |\epsilon _1| ^2 + |\epsilon _2|
^2 + ... \ll 1 \).
The hadronic loop contributions allow the bare states ($s \overline s$ in
this example) to communicate with all hadronic channels permitted by quantum
numbers, and this enables the meson to decay, its lifetime being inversely
proportional to the width $\Gamma$.
However, in this case the switching on of interactions produces a relatively 
tiny effect 
and the physical $\phi$ is still overwhelmingly an $s \overline s$ state.
For this reason the naive quark model works very well; so from the observed
hadron we can easily infer its quark structure.
A similar picture works for the tensors.

\noindent
For scalar mesons the situation is different and the one--to--one 
correspondence between the observed scalar mesons and their underlying quark 
content is distorted by dynamical effects. This is because they couple 
strongly to more than one meson-meson channel, creating overlapping and 
interfering resonance structures. 
Furthermore, since the interactions are $S$--waves, the opening of each 
threshold  
produces a more dramatic $s$-dependence in the propagator. 
At each threshold, there is a centrifugal barrier factor of $k^L$, where $k$ 
is the appropriate c.m. $3$-momentum of the decaying products and $L$ their 
relative orbital angular momentum. 
This means that the thresholds for higher spin states open more smoothly.  

\noindent
Let us now go into more detail about the model starting, 
for simplicity, from the case in which only one underlying bare state has 
to be considered.
This is the case, for example, for the $I=1/2$ and $I=1$ sectors, the seeds
of  which are $s\overline n$ and $n\overline n$ respectively.
We define a vacuum polarization function $\Pi (s)$ which 
accounts for all the possible two pseudoscalar 
loop contributions to the propagator $P(s)$ \cite{torn}. 
Referring to the pictorial
representation of Eq.~(\ref{fullprop}) in Fig. \ref{prop}, 
we can easily write 
$\Pi$'s imaginary part:
\be
{\rm Im} \Pi (s) \,=\, - \sum _{i} G_i ^2 (s) \,=\,  
- \sum _{i} g_i ^2 \, \frac{k_i(s)}{\sqrt s} \, (s -s_{A,i}) \,  F_i ^2(s)
\, \theta (s-s_{th,i})
\label{Im}
\ee
where the index $i$ runs over the pseudoscalar channels and the
$k_i$'s are the c.m. momenta of the two intermediate pseudoscalars. 
The $g_i$'s are the $SU(3)$ flavour couplings connecting the bare state to
the two--pseudoscalar loop: \(g_i = \gamma \gamma _i\), see 
Refs.~\cite{tornold,torn} for more details.  
The terms $(s-s_{A,i})$ give the Adler zeros, required for $S$--waves by 
chiral dynamics.
$F_i(s)$ are the form factors, which take into account the fact that the 
interaction is not pointlike but has a spatial extension. 

\noindent
Since the vacuum polarization function, $\Pi(s)$, is an analytic function, 
its real part can be
deduced from the imaginary part by making use of a dispersion relation 
\be
{\rm Re} \, \Pi (s) \,=\, \frac{1}{\pi} P \int _{s_{th,1}} ^\infty ds' \; 
\frac{{\rm Im} \Pi (s')}{s'-s}\;.
\label{Re}
\ee 
No subtraction is needed, since the form factors are built in such a way
that they decrease fast enough when $|s| \to \infty$.

\noindent 
At this point, we can write the propagator in terms of the vacuum
polarization function:
\be
P(s) \,=\, \frac{1}{m_0 ^2 + \Pi (s) - s } 
\label{vacuum-P}
\ee
The mass and the width of the decaying hadron are determined, in a process
independent way, by the pole of the propagator. Consequently, in order to
find the pole position, we have to 
continue Eqs.~(\ref{Im}, \ref{Re}, \ref{vacuum-P}) into the complex 
$s$--plane onto the appropriate unphysical sheets.

\noindent
The contribution of this resonance pole to the $i \to j$ amplitude
is then 
\be\,\,
A_{ij} (s) \,=\, \frac{G_i(s)\,G_j(s)}{m_0 ^2 + \Pi (s) - s} \; . 
\label{A}
\ee
which respects the unitarity requirement,
\(
A-A^{\dagger} = 2i \, AA^{\dagger}\).
As a consequence, for each elastic channel we can define a resonant 
phase-shift by
\be
A_{ii}(s)\,=\, \frac{1}{2i}\,\left( \eta_i\ e^{2i\delta_i}\,-\, 1\right)\; .
\ee

\noindent 
The amplitude $A_{ij}$ in Eq.~(\ref{A}) represents a generalization of the 
well-known Breit-Wigner formula, at least in the neighbourhood of the pole. 
This is readily seen by writing the vacuum polarization function in terms 
of its real and imaginary parts, then
\be 
A_{ij} (s) \,=\,  
\frac{G_i(s)\,G_j(s)}{m_0 ^2 + {\rm Re}\Pi (s) -s + i{\rm Im} \Pi (s)}
\,=\, 
\frac{m_{R}[\Gamma _i(s)\Gamma _j(s)]^{1/2}}{m^2(s) -s -i\,m_{R} 
\Gamma _{tot} (s)}\;,
\label{A-BW}
\ee
having identified 
\be
\Gamma _{tot}(s)= -\frac{{\rm Im} \Pi (s)}{m_{R}}
= \sum_i \, \Gamma_i(s)\;,
\ee
where 
\be
\Gamma _i(s)=\frac{G_i^2(s)}{m_{R}}\;,
\label{width}
\ee
and
\be
m^2(s)=m_0 ^2 + {\rm Re}\Pi (s)\;.
\ee

\noindent
Here $m^2(s)$ is the {\it running squared mass}, given by the sum of the bare
mass squared and the real part
of the vacuum polarization function ${\rm Re}\Pi (s)$, which is responsible for
the {\it mass shift}.
The imaginary part of the vacuum polarization function 
${\rm Im}\Pi (s)$ is directly proportional to the width of the state.
The mass shift function ${\rm Re}\Pi (s)$ is generally negative and is
approximately constant only in the energy regions far from any threshold.  For
$S$--waves the $s$--dependence becomes crucially important nearby 
thresholds, since ${\rm Re}\Pi (s)$ has square root cusps at 
the opening of each of them.

\noindent
Though the only correct way to calculate the mass of a particle is to find
the position of the propagator pole in the complex $s$--plane, it is
useful to define another quantity, again obtainable from the propagator, 
which we will call the Breit--Wigner mass 
$m_{BW}$. It corresponds to the intersection of the running mass $m^2(s)$ 
with the curve $s$, {\it i.e.} the particular value of $s$ where the function 
$(m_0^2+{\rm Re} \Pi(s) -s)$ vanishes, with $s$ wholly real:
\be
m^2_{BW}=m_0^2+{\rm Re} \Pi(m^2_{BW})\;.
\ee
This gives a rough estimate of what the physical mass is.
As a matter of fact, moving into the complex plane the real coordinate of 
the pole position can change considerably if thresholds to strongly coupled 
channels are located nearby,  
and when the dynamics are particularly complicated. 
As  $m_{BW}$ of Eq.~(13) is defined as the energy at which the 
function $m^2(s)$ is equal to $s$, this is a point that we refer to 
as a {\it crossing} for reasons that will become clear in Fig.~\ref{am}. 
Clearly
at $\sqrt{s} = m_{BW}$, the amplitudes $A_{ij}$ of Eq.~(9) become purely 
imaginary. 
It is this simple fact and its physical consequences that are the 
theme of this paper.

\newpage

\section{$I=1$ sector and the role of the Wigner condition}

\noindent
We now turn our attention to the issue of accomodating all the scalar meson 
states (which experiments deliver) in either one or more quark model 
multiplets. In Ref.~\cite{torn} Tornqvist finds a scalar nonet composed of the 
$K_0^*(1430)$, the $a_0(980)$, the $f_0(980)$ and the $f_0(1370)$. 
Furthermore, in Ref.~\cite{roos}, one extra low energy pole is 
found, which the authors identify with the much discussed broad $\sigma$ 
meson, called $f_0(400-1200)$ in the Particle Data Tables~\cite{PDG}. 
Nevertheless, this 
study leaves out the $a_0(1430)$, for instance, for which Crystal 
Barrel~\cite{CrBarr} finds clear evidence.

\noindent
Consequently, we begin by examining the $I=1$ sector, since this is 
relatively simpler than the others. We ask: 
can we \lq\lq generate\rq\rq ~one or more extra physical states in the same 
sector, in the framework of our simple model, by starting from only one 
$\overline n n$ bare seed ?    

%
\begin{figure}[t]
\begin{center}
\mbox{~\epsfig{file=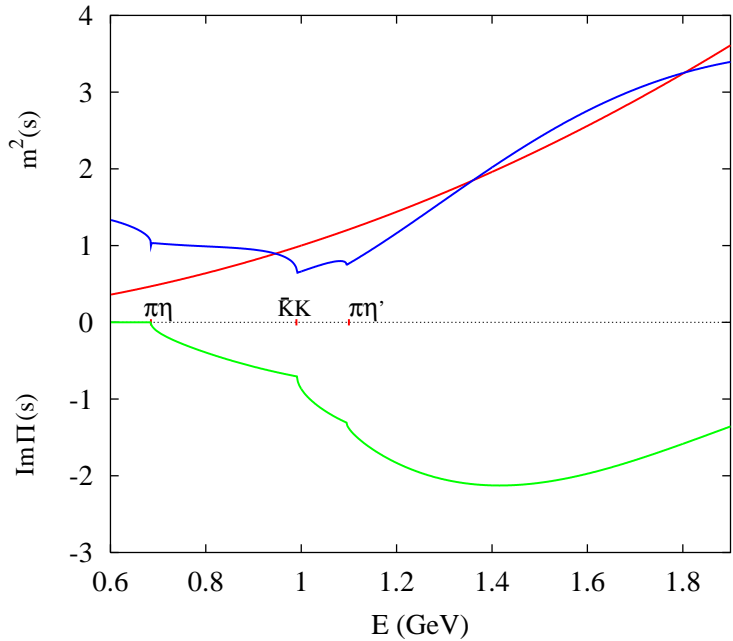,angle=-90,width=10cm}}
\caption{\label{am} \footnotesize{
The curves $m^2(s)$,  $s$ and $\rm Im\Pi (s)$ as functions of the energy 
$E\ =\ \sqrt{s}$, for the sector $I=1$.
Each intersection between $m^2(s)$ and $s$ is referred to 
as a {\it crossing}.
The first of these is 
situated at approximately $915$ MeV, between the 
$\pi \eta$ and the $K\overline K$ threshold. 
The second crossing, well above the 
$\pi \eta ^{\prime}$ threshold, is at $E = 1430$ MeV. The third 
intersection, at $1.82$ GeV, is a non-physical state according to 
the Wigner condition.
}}
\end{center}
\end{figure}
%

\noindent
By increasing the overall coupling $\gamma$ and the bare mass of the 
$\overline n n$ seed ($\gamma = 1.53$, $m_0 = 1.620$ GeV, as opposed to 
$\gamma = 1.13$, $m_0 = 1.420$ GeV used in Ref.~\cite{torn}), we find it is 
possible to obtain a scenario in which more than one intersection between the 
mass function $m^2(s)$ and the $s$ curve appear in the mass plot, 
as shown in Fig.~\ref{am} (upper half).
The first crossing, situated at approximately $915$ MeV, between the 
$\pi \eta$ and the 
$K\overline K$ threshold, corresponds to a state which can be identified with 
the $a_0(980)$. We treat the charged and neutral kaons as degenerate in mass
and so neglect the possibility of isospin mixing between $I=1$ and $I=0$ 
states \cite{achasov}.
The second crossing, occurring well above the 
$\pi \eta ^{\prime}$ threshold, at $m_{BW} = 1430$ MeV, is again a 
physical state and has the right 
features to represent the Crystal Barrel relatively broad $a_0(1450)$; 
the third intersection occurs at $E=1.82$ GeV.
In the lower half of Fig.~\ref{am} we plot the curve $\rm Im \Pi (s)$, which 
shows 
how the second state which we identify with the $a_0(1450)$ is much broader 
than the $a_0(980)$. 

\noindent To know which state is physical we refer to the 
Wigner condition~\cite{Wigner}. 
This condition follows from the principle of causality, {\it i.e.} the 
requirement that the scattered wave does not leave the scatterer before the 
incident wave has reached it.
In the present context, Wigner's theorem limits the rate of fall of 
the phase-shift,
$\delta_{ij}$. A physical resonance cannot occur if the phase shift falls 
through $90^o$. In fact, such a resonance would have 
a negative width and correspond to a pole on the upper half plane in the 
physical sheet. It would therefore represent a non-causal, exponentially 
increasing state. 
Since it is not possible to have a state with a lifetime greater than 
the period of scattering, Wigner's condition requires
\be
\frac{d\delta}{dk} \ge -\frac{1}{m_{\pi}} \frac{\sqrt{s}}{2k}.
\label{Wigner}
\ee 
where the phase-shift $\delta$ refers to the same channel as 
the c.m. 3-momentum $k$ and is defined by Eq.~(8).
The Wigner condition is particularly useful in the inelastic region, where 
resonances do not necessarily correspond to $\delta=90^o$, but more generally 
occur when $\delta=n\pi/2$ and the real part of the scattering 
amplitude is zero.  
%
\begin{figure}[t]
\[\begin{array}{ll} \hspace*{-1.0cm}
\mbox{~\epsfig{file=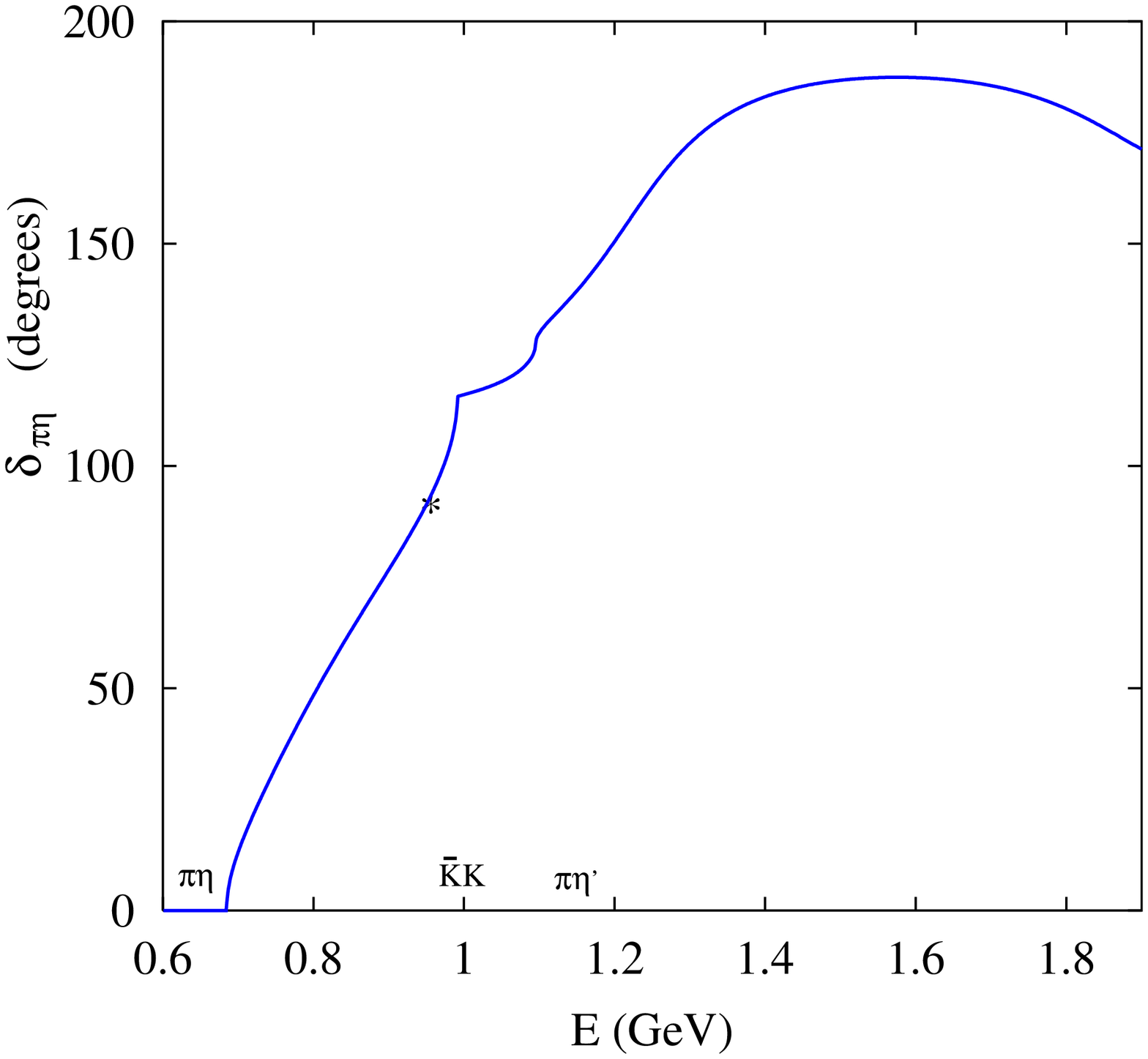,angle=-90,width=8cm}} 
& 
\mbox{~\epsfig{file=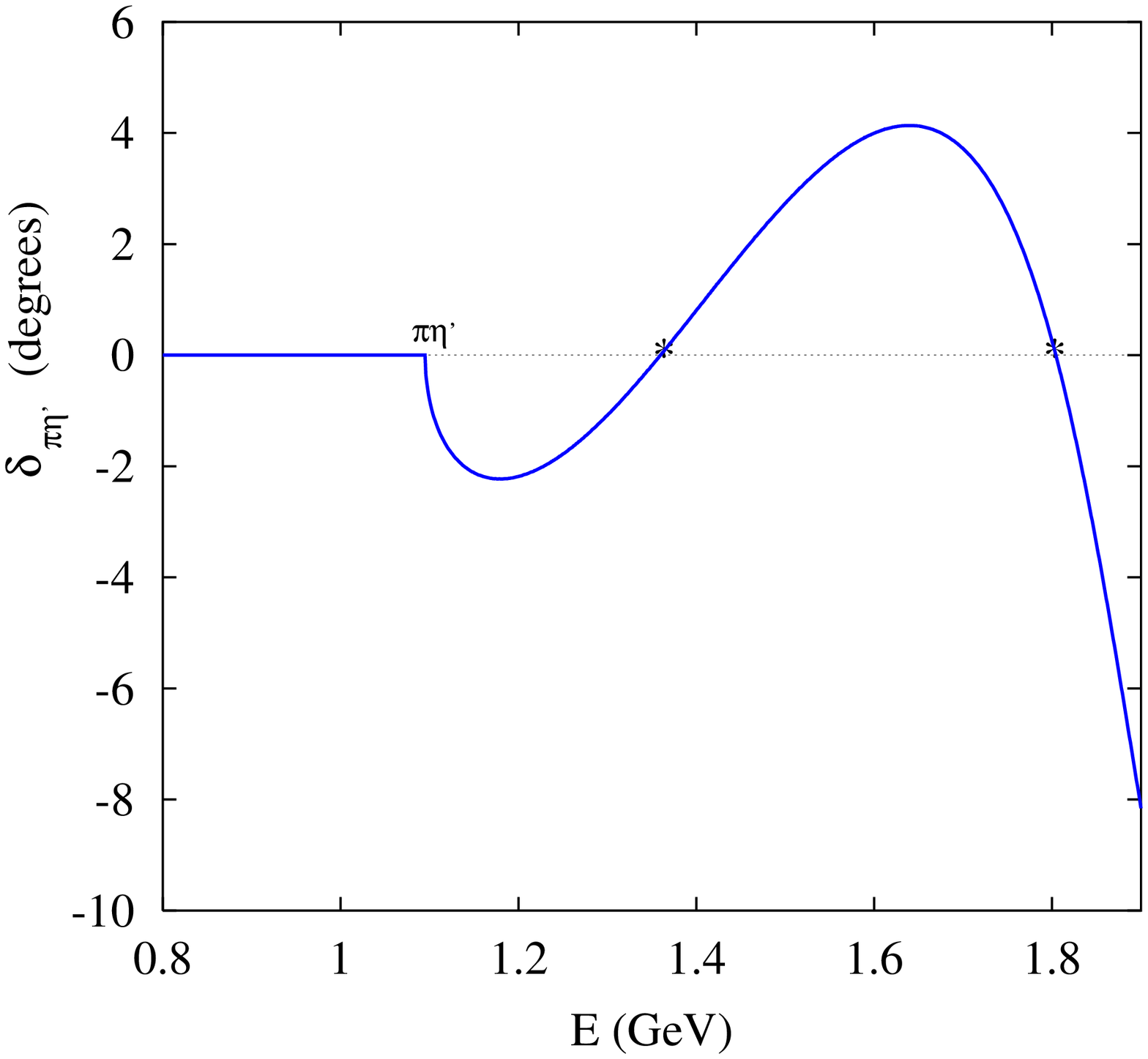,angle=-90,width=8cm}} 
\end{array}\]
\caption{\label{aph} \footnotesize{
The phase $\delta_{\pi\eta}$ reaches $90^o$ at  
$E=915$ MeV, corresponding to the first physical state identified with the 
$a_0(980)$ (the point is labelled by a star on the left hand plot).  
The phase $\delta_{\pi\eta ^{\prime}}$ reaches zero twice (the two points 
are labelled on the right hand plot), the first time 
raising from negative to positive values and giving the second physical 
$a_0$ state, the second time quickly dropping from positive to negative 
values, and delivering the unphysical third state, as confirmed by the 
Wigner condition.
}}
\end{figure}
%
\begin{figure}[t]
\begin{center}
\mbox{~\epsfig{file=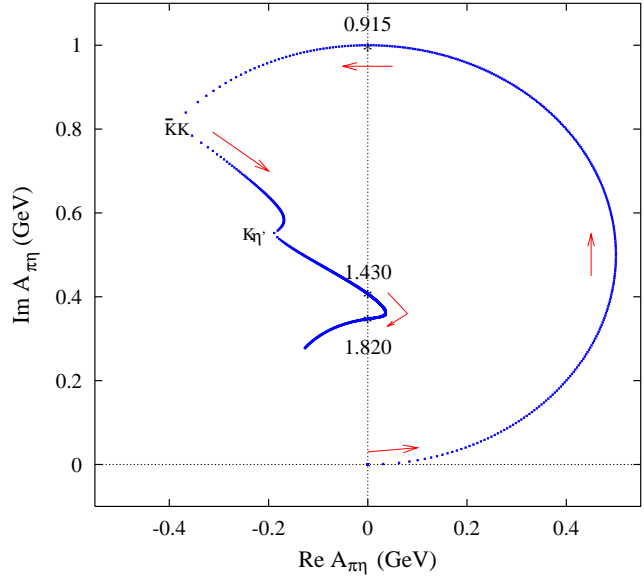,angle=-90,width=9cm}}
\caption{\label{aarg}\footnotesize{
Argand plot of the amplitude 
$A_{\pi\eta}$: in the elastic region, Im$A_{\pi\eta}$ raises quickly 
anti-clockwise and Re$A_{\pi\eta}$ 
reaches zero at $E=915$ MeV, the lighter physical $a_0$. The second and third 
states occur above $\pi\eta ^{\prime}$ threshold, as indicated by the two 
stars on the plot. Notice that the heavy unphysical state is situated 
immediately after an abrupt change of direction in the curve Im$A_{\pi\eta}$.}}
\end{center}
\end{figure}

\noindent
Fig.~\ref{aph} shows the phases, $\delta_{\pi\eta}$ and 
$\delta_{\pi\eta ^{\prime}}$, which tell us about the characteristics of the 
states we find: 
$\delta _{\pi\eta}$ goes through $90^o$ at $E=915$ MeV, in correspondence with 
the first physical resonance, 
the lighter $a_0$. Instead, $\delta_{\pi\eta ^{\prime}}$ reaches zero twice: 
the first time 
rising from negative to positive values and giving the second physical 
$a_0$ state, the $a_0(1450)$; the second time quickly dropping from positive 
to negative values: the third crossing at 1820 MeV is 
unphysical as given by Eq.(\ref{Wigner}).

\noindent Finally, Fig.~\ref{aarg} shows the Argand plot of the amplitude 
$A_{\pi\eta}$: in the elastic region, Im$A_{\pi\eta}$ rises quickly 
anti-clockwise and Re$A_{\pi\eta}$ 
reaches zero at $E=915$ MeV, the lighter physical $a_0$. The second and third 
states occur above $\pi\eta ^{\prime}$ threshold, as indicated by the two 
stars on the plot. Notice that the heavy unphysical state is situated 
immediately after an abrupt change of direction in the curve 
Im$A_{\pi\eta}$.

\noindent Our conclusions about yet heavier states are not reliable in the 
present model, since higher thresholds are not included which may well 
affect the picture at higher energies dramatically.
Nevertheless just including the lightest two pseudoscalar channels 
we can definitely 
say that a scenario in which not only one but a 
series of scalar physical states with I=1 can easily be achieved,  by
 fine tuning key free parameters of the model. 

\noindent
Generating more than one 
physical state with the same quantum number starting from one seed brings
us closer to the picture emerging from experiment.  
Will this be the case for the other sectors as well ? And to what extent is
 the adjustment of the overall coupling and mass parameters permitted 
within this model? 
That is what we discuss in the following Sections.

\section{The $I=1/2$ sector}

\begin{figure}[t]
\begin{center}
\mbox{~\epsfig{file=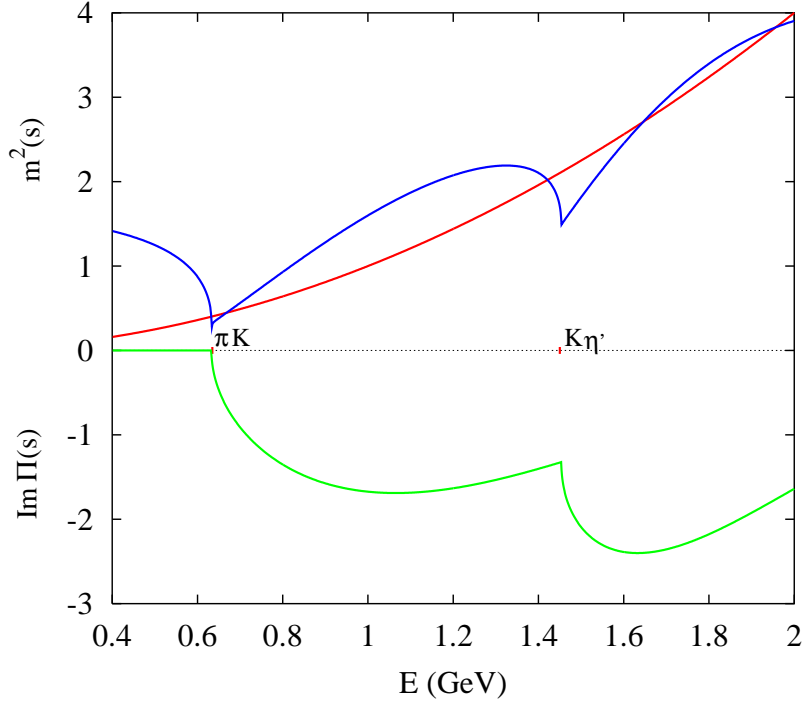,angle=-90,width=11cm}}
\caption{\label{km}\footnotesize{
The curves $m^2(s)$,  $s\ =\ E^2$ and $\rm Im\Pi (s)$ as functions of the 
energy $E$. The first crossing point is located  
{\it below} the $K\pi$ threshold and corresponds to a pole on the real axis 
({\it i.e.} a non-decaying state). The second crossing is again an unphysical 
state because it does not satisfy the Wigner condition, as clearly shown by 
the phase shift and the Argand plot in Figs.~\ref{kph} and \ref{karg}. 
On the contrary, the third intersection point corresponds to an unphysical 
state. Finally the last two crossings correspond to a non-physical and a 
heavier and broader physical state respectively. 
}}
\end{center}
\end{figure}

In an analogous manner to the  $I=1$ case, we now want to examine 
the possibility of dynamically generating more physical states with the same 
quantum numbers from only one bare seed,  $n\overline s$ in this particular 
case. 
Here the main issue is to investigate whether it is possible to generate the 
light  $I=1/2$ state called $\kappa$, advocated for example in 
Refs.~\cite{oller-oset, schecter, vanbev}.\\
The situation for the $I=1/2$ sector is rather different from the $I=1$ case. 
Here there are only two relevant thresholds and their positions and coupling 
strengths are such that 
the shape of the mass function curve, $m^2(s)$, is very rigid and varying the 
parameters changes this little.
Nevertheless, by using the same changes we used for the $I=1$ sector 
($\gamma = 1.53$, $m_0 = 1.620$ GeV, with $s_{A,\pi K}= -1.0$ GeV$^2$)  
we obtain the plot shown in Fig~\ref{km}.

\begin{figure}[p]  
\[\begin{array}{ll} \hspace*{-1.0cm}
\mbox{~\includegraphics[width=6.0cm,angle=-90]{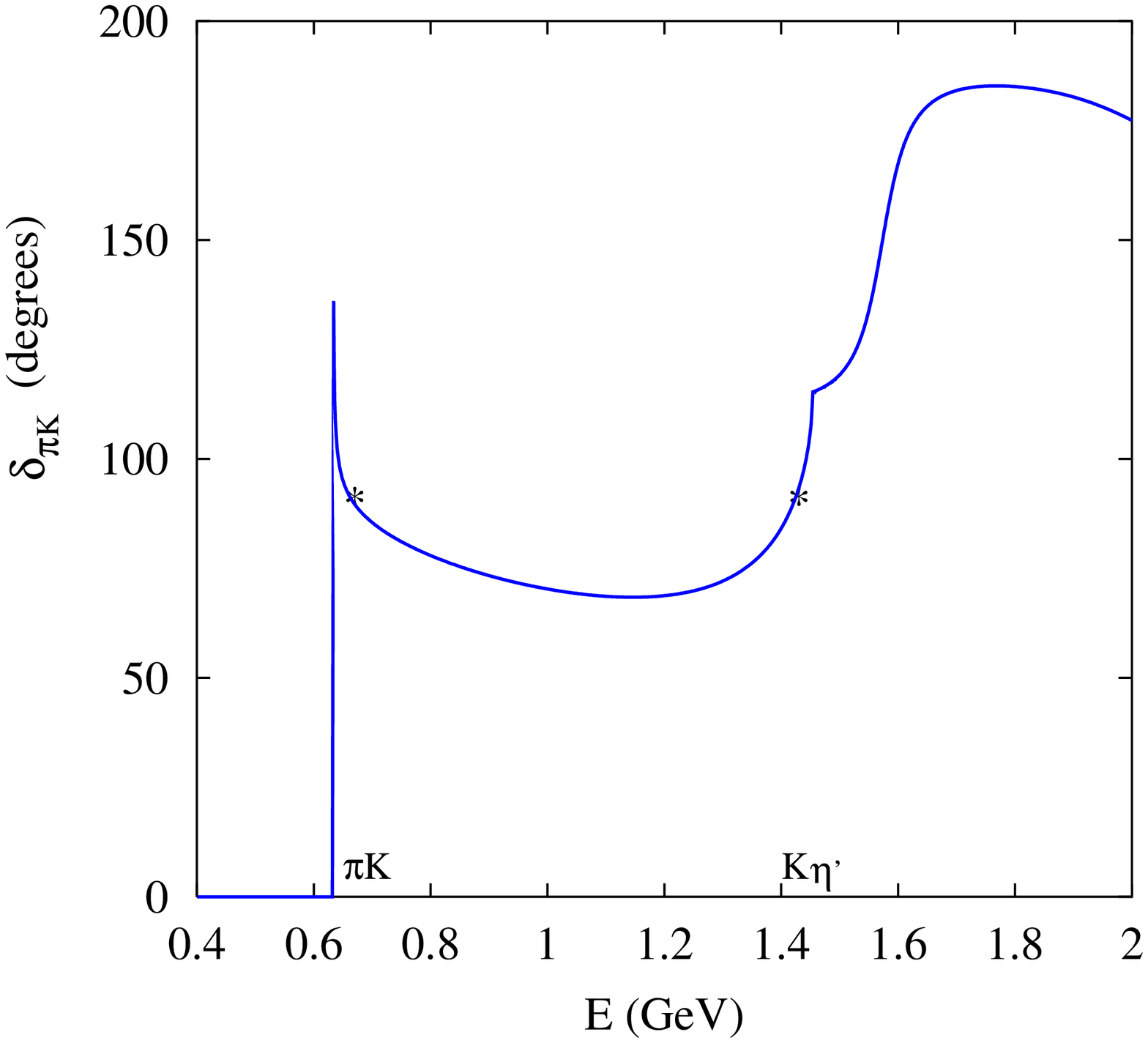}} 
& 
\mbox{~\includegraphics[width=6.0cm,angle=-90]{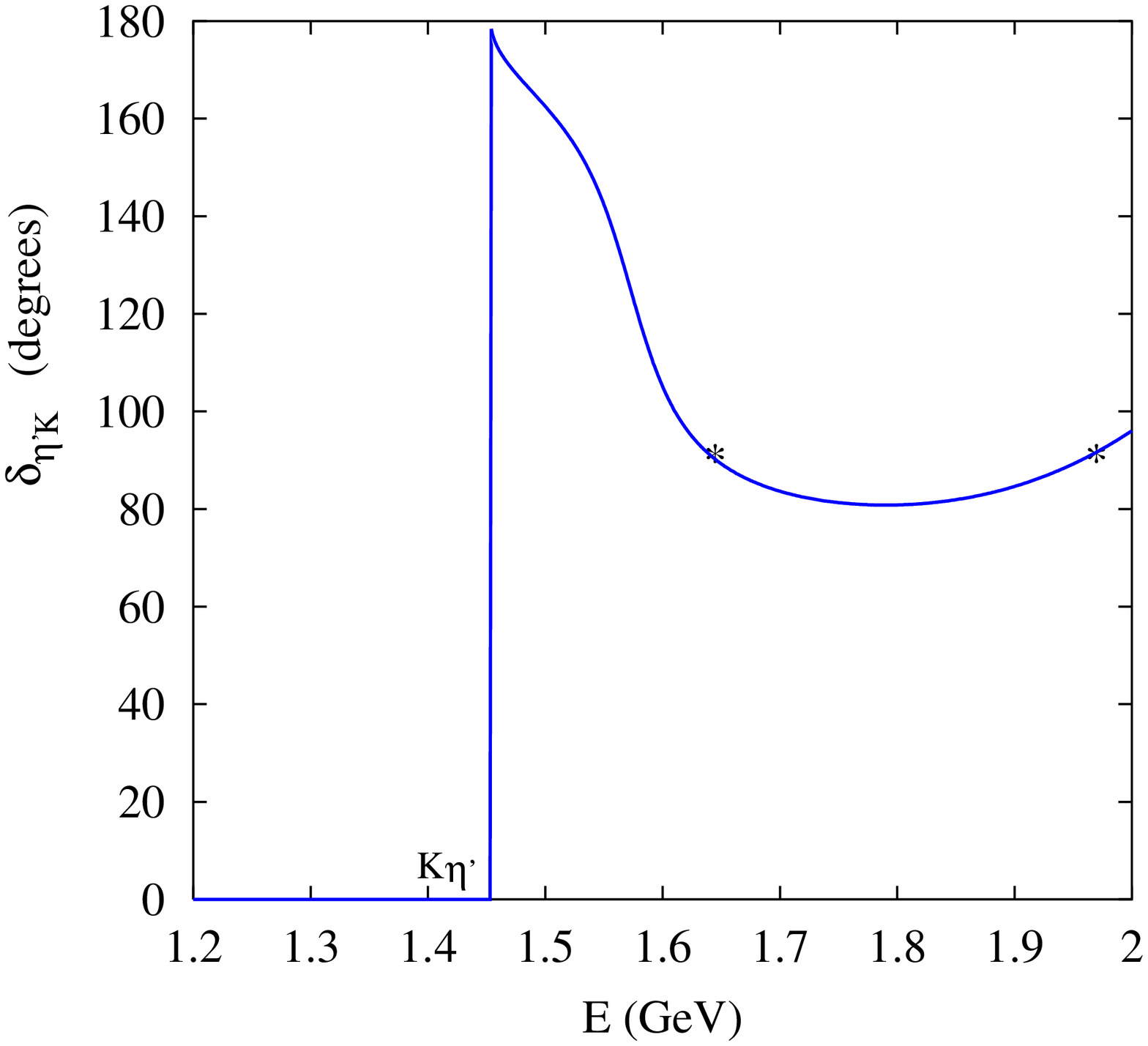}}
\end{array}\]
\caption{\label{kph} \footnotesize{
The phase $\delta_{K\pi}$ drops extremely quickly through 
$90^o$ at $E=670$ MeV, in correspondence with
 the unphysical state generated by  
the second crossing point in Fig.~\ref{km}. It then rises 
through $90^o$ at $E=1420$ MeV, signalling the presence of a physical state, 
 (the two points are labelled on the left hand plot).
The phase $\delta_{K\eta^{\prime}}$ goes through $90^o$ twice 
(the two points  are each labelled by a star on the right hand plot), 
the first 
time decreasing quickly from $180^o$ at $K\eta ^{\prime}$ threshold, giving 
the third unphysical state, the second time while increasing again to larger  
values, and delivering the last physical state, as confirmed by the 
Wigner condition.}}
\begin{center}
\includegraphics[width=7cm,angle=-90]{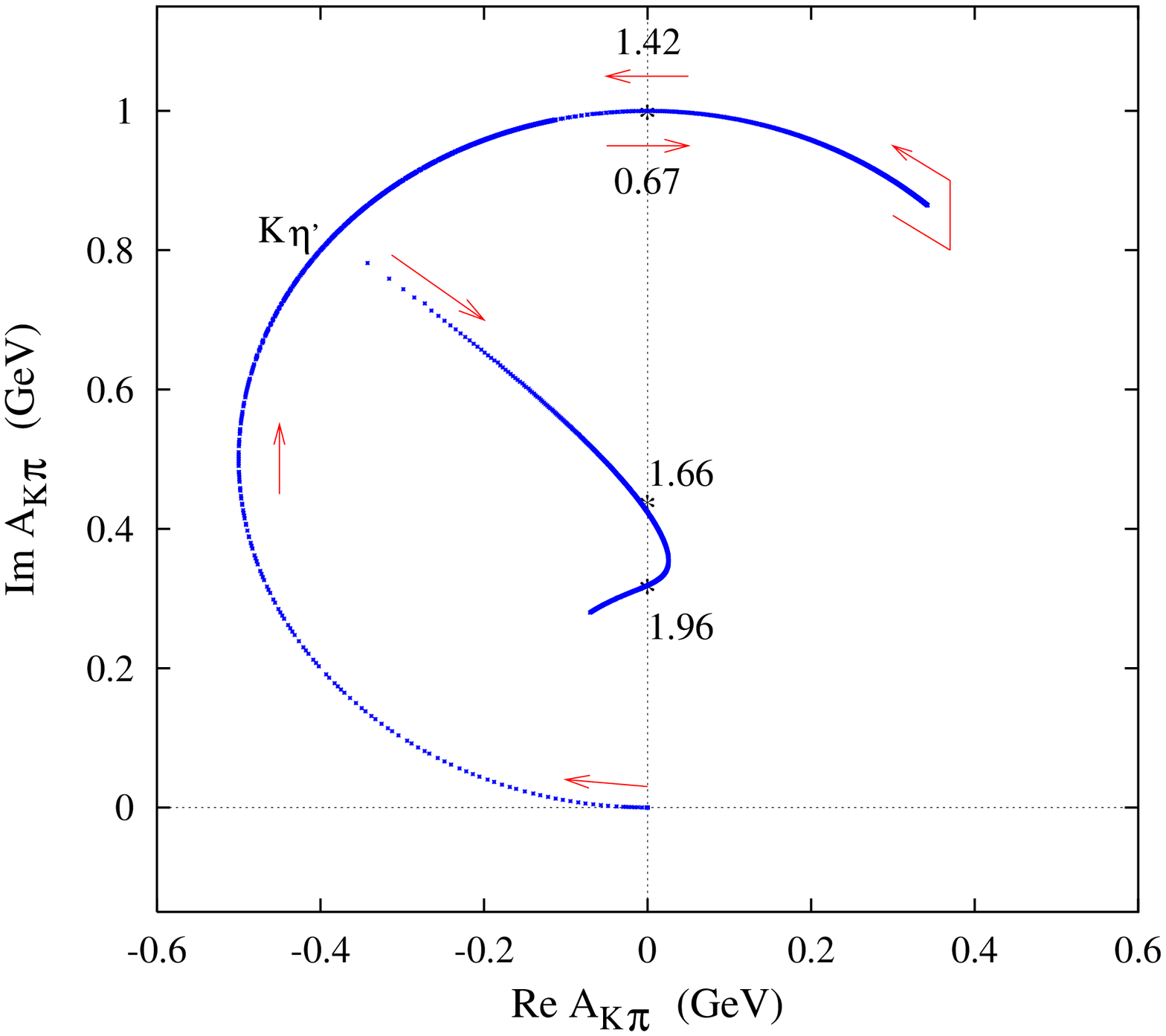}
\end{center}
\caption{\label{karg}\footnotesize{
Argand plot of the amplitude 
$A_{K\pi}$: in the elastic region, Im$A_{K\pi}$ raises quickly clockwise and 
Re$A_{K\pi}$ reaches zero at $E=670$ MeV, corresponding to the first 
unphysical state. It then goes back anti-clockwise and Re$A_{K\pi}$ reaches 
zero again at $E=1420$ MeV, signalling the presence of the first physical 
state, the $K_0^*(1430)$ (indicated on the plot) by a star.
The following unphysical and physical states, with BW masses of $1.66$ 
and $1.96$ GeV respectively, are situated in the inelastic region, well over 
the $K\eta ^{\prime}$ threshold, and are also 
indicated on the plot by little stars.
}}
\end{figure}

\noindent
The first interesting observation is that the first point of intersection  
between the mass function $m^2(s)$ and the curve $s$, is always bound to be 
{\it below} the $K\pi$ threshold and it corresponds to a pole on the real 
axis, 
{\it i.e.} to a state with zero width: in fact, the cusp signals the precise 
location of the threshold itself, so that varying the coupling strength only 
slightly alters the position of the crossing point, but never allows it to 
move above the threshold. 
%
%
%
The second intersection point is not a physical state 
since it violates the Wigner condition, as is clearly shown by the 
phase shift and the Argand plot in Figs.~\ref{kph} and \ref{karg}. 
Notice once again that the conclusion that 
this state is not physical does not 
depend on the strength of the coupling to the $K\pi$ channel: varying this  
only produces a tiny shift in the position of the crossing point, and does 
not alter the characteristics of the state. 
Thus within this model having just one  
$n\overline s$ bare seed, it turns out to be impossible to generate a 
physical $\kappa$-like state.
In contrast, the third intersection point corresponds to a physical state
 which can easily be interpreted as the $K_0^*(1430)$, already found in 
Tornqvist results of Ref.~\cite{torn}.
It is interesting to notice that the same kind of picture would emerge when 
using the parameterization proposed in Ref.~\cite{roos}: the first crossing 
point is again below threshold and the second one corresponds to a state 
which, violating the Wigner condition, is unphysical. 
Indeed, only the third intersection point is a physical one.

\noindent
Moving to higher energies, we find two further crossing points between the 
$s=E^2$ curve and the mass function $m(s)^2$: according to the Wigner 
condition, only the last one, situated at $E=1.96$ GeV, corresponds to a 
physical state.  
As we mentioned in the previous Section, results concerning states with heavy 
masses are not completely reliable since the present model does not include 
higher meson-meson thresholds. Nevertheless, it is quite interesting to notice 
that the last physical state predicted by our calculation could be 
identified with the $K_0^*(1950)$ reported by the PDG group (see table on 
p. 51 of Ref.~\cite{PDG}).

\noindent
Compared with other results available in the literature, 
as far as the $I=1/2$ and $I=1$ sectors are concerned, ours are 
similar to those presented in the work of  
Minkowski and Ochs \cite{ochs}, where they claim the
existence of two complete nonets in the light scalar meson spectrum, 
the first one
characterised by the $K_0^*(1430)$ and the $a_0(980)$, and the second one
by the $K_0^*(1950)$ and the $a_0(1450)$. To the isoscalar sector they
assign the $f_0(980)$, $f_0(1500)$ and the $f_0(1720)$, $f_0(2020)$ 
respectively, with singlet-octet mixing angle (much as in the pseudoscalar 
nonet), having
dismissed the $\sigma$ and the $f_0(1370)$.\\
In contrast, van Beveren et al. \cite{vanbev} find two lower mass nonets, 
given by 
\[ \kappa , \; a_0(980), \; f_0(980), \; \sigma \]  
\[K_0^*(1430),\; a_0(1450), \; f_0(1370), \; f_0(1500)\;,\]
whereas Shakin and collaborators \cite{shakin} predict the existence of  up 
to three scalar 
nonets, characterized by 
\[a_0(980), \; K_0^*(1430)\]
\[a_0(1450),\; K_0^*(1730)\]
\[a_0(1857),\; K_0^*(1950)\]
and keep the $\sigma$ and $\kappa$ mesons out, using their particular 
definition of ``dynamically generated states'' with no right to be classified 
into a nonet structure. Moreover the $a_0(1857)$ and the $K_0^*(1730)$ are 
resonances which appear in their calculation but are not confirmed by 
experiment yet. 

\noindent
To complete our picture we now move to the  $I=0$ sector.

%

\section{The $I=0$ sector}
    
This is the most complex and delicate sector, and drawing any kind of 
conclusion is particularly difficult: let's see why. As discussed in 
the previous Sections,
in the $I=1$ and $I=1/2$ sectors physical states originate from one 
``bare seed'' only, either $n\overline n$ or $n\overline s$, and are then 
``dressed'' by coupling to meson-meson channels. In contrast, in the $I=0$   
case we already have two possible ``bare seeds'' to start with,   
$n\overline n$ and $s\overline s$, which not only get dressed but in fact do  
{\it mix} through hadronic loops. And mixing entangles the situation to such 
an extent that the one--to--one correspondence between the phase 
shift behaviour and the occurrence of physical states is lost: then   
the Wigner condition cannot be used to sort the physical states from 
the unphysical ones.
Moreover, the couplings of the isoscalar bare seeds to the relative 
meson-meson channels, $\pi\pi$, $K\overline K$, $\eta\eta$ etc ... 
are not calculated in a standard and unique way: 
there are several calculations in the literature, each of which is based on 
different (but in our opinion equally good or bad) assumptions 
(see for example Refs.~\cite{torn} and \cite{vanbev}).
So one has to make a choice of coupling scheme as well, and unfortunately 
the outcome strongly depends on this choice.

\noindent
If we simply use for the parameters $\gamma$ and $m_0$ the same 
values as in Sections 3 and 4 and the coupling scheme as given  
by Tornqvist in Ref.~\cite{torn} (with the $\eta\eta^{\prime}$ threshold 
coupling enhanced by a factor $\beta = 1.6$) or by van Beveren {\it et al.} 
\cite{vanbev}, 
we find scenarios in which multiple crossings do occur and many isoscalar 
states are generated. 
Unfortunately, we cannot tell them apart, because the phase shift behaviour 
does not help, especially in the region around and above $1$ GeV, where the 
mixing is maximal and the overlapping among resonances is crucial. 
And we cannot say 
which of them are physical states either, because the Wigner condition 
cannot be applied.
Moreover, as we anticipated, the position and features of the crossing 
points are very strongly model dependent, and very different scenarios can 
be created by varying the coupling scheme, so that no robust model 
independent conclusion 
can be reached in this sector. For instance, 
if we apply purely $SU(3)$ pseudoscalar-pseudoscalar couplings, as presented 
in Refs.~\cite{torn, tornold} without  
enhancing the $\eta\eta^\prime$ threshold, only two resonant states are 
found.

\noindent  
Nevertheless, it is a fact that a number 
of states larger than two {\it can} be created starting from just two bare 
seeds. For certain parameter 
configurations and with certain coupling schemes, the 
$I=0$ experimental candidates can be accounted for. This shows  
that the dynamical generation of physical states {\it is} a possible solution 
to the problem of accounting for more scalar mesons than can fit in one nonet. 
Notice, though, that this is a ``democratic'' model, in which we cannot 
distinguish  which are the intrinsic (or pre-existing) states as opposed to 
the dynamically generated ones (a discussion about this issue was raised 
by van Beveren {\it et al.} in a comment \cite{vanbev-comm} on a paper by 
Shakin and Wang \cite{sw}).  

\section{Scattering amplitudes}

To what extent are we free to change the parameters of the model to allow the
double crossing to occur that generates the $a_0(1430)$ as well as the 
$a_0(980)$? This question is addressed by considering the relationship of 
the propagators in Eqs.~(1,2,6,7) to the corresponding physical scattering 
amplitudes.
As noted in  
Section 2, knowing the propagator of $s-$channel resonances
determines a key part of the scattering amplitudes, Eqs.~(7,9).
 For Tornqvist and Roos, this amplitude $A_{ij}$ is {\it all} there is to the 
hadronic scattering amplitude. However, $s-$channel dynamics is not all that 
controls the scattering. So while the amplitude $A$ defined in Eq. (\ref{A}) 
respects  unitarity, it is not the most general amplitude that achieves this, 
having its numerator and denominator related by the same couplings $G_i$ 
(see Eq. (\ref{Im})).
Knowing the structure of the propagator $P(s)$, Eq.~(\ref{vacuum-P}), 
we can write in complete generality the full scattering amplitude as
\be
T(s) = \frac{N(s)}{m^2(s)-s+\Pi(s)}\;,
\label{general}
\ee
where we drop the channel labels $ij$. The numerator $N(s)$ is an 
unknown complex function of the variable $s$, 
which can be re--expressed in terms of its modulus and phase as
\be
N(s)=|N(s)|\,e^{i\alpha(s)}\;.
\ee
Imposing both elastic and inelastic unitarity one finds that
\be
|N(s)| = [m^2(s)-s+{\rm Re}\Pi(s)] \sin \alpha (s) - {\rm Im}\Pi(s) 
e^{i\,\alpha (s)} \;, 
\ee
which allows the most general hadronic amplitude to be written as
\be
T(s) \,=\, A(s) \,  e ^{\,2 i \, \alpha (s)} + \sin \alpha (s) \;
e^{\,i \, \alpha (s)}
\label{R}
\ee
where $\alpha (s)$ is an unknown function of $s$ real along the right 
hand cut,
and the second term in Eq.~(\ref{R}) can be regarded as a background 
contribution. It is important to note that if the phase-shift of the amplitude
$A$ is $\omega$, then the phase-shift of the full amplitude $T$ is
$\delta = \omega + \alpha$.\\
The model of Tornqvist and Roos is to set $\alpha=0$ everywhere.

\noindent
In Ref.~\cite{thesis} we have improved on Tornqvist's study by 
performing new fits to experimental data, based on the general 
amplitude $T(s)$ rather than the pole dominated amplitude $A(s)$. 
From this we conclude that very good 
solutions can be obtained by using relatively small values for the 
parameter $\alpha$ (we chose a constant value of $15^o$ as an example) and 
varying the $m_0$ and $\gamma$ parameters by a few hundred MeV. For instance,
to fit $\pi K$ scattering data in terms of the pure pole amplitude $A$, 
Tornqvist  requires the position of the Adler zero
to be at $s_{A,\pi K} = -0.42$ GeV$^2$ far from its position in chiral 
perturbation theory (current algebra predicts 
$(m_K-m_{\pi})^2 \le s_{A,{\pi K}} \le (m_K+m_{\pi})^2$ with an average value 
of $0.24$ GeV$^2$). 
In contrast, by a simple choice of $\alpha$ of about 
$15^0$ we can fit these same data with the full hadronic amplitude $T$ which 
has the Adler zero at the position required by chiral dynamics.  

\noindent 
This puts our attempts to generate extra states with the same quantum 
numbers by varying the free parameters of the model on more solid ground. 
We know that small changes in the value of $m_0$ and $\gamma$ can give 
equally good fits using the full hadronic amplitudes, $T$.
This gives us confidence in our general conclusions.
The fact that there is more to a scattering process
than $s-$channel dynamics means that fitting data along the real axis cannot
accurately determine the true pole position of a broad state without an 
analytic continuation, or a very specific model. This  casts doubt on
the determination of the position of the $\sigma-$pole
by Tornqvist and Roos. 
Their fit to $\pi\pi$ data in terms of the amplitude $A$ gives quite 
different parameters than using the full amplitude $T$ of Eq.~(\ref{R}).
That there is more to dynamics than $s$-channel resonances has been noted 
by Isgur and Speth in this same context \cite{isgur-speth}.

\section {Conclusions} 

The present work focusses on the study of the 
$I=1$ and $I=1/2$ sector of the light scalar meson spectroscopy.  Previous
papers from Tornqvist and Roos~\cite{torn,roos} seemed to suggest that
using a simple model based on the hadronic ``dressing'' of bare seeds, one
could generate more than one, possibly a whole family of mesons, with the
same quantum numbers, starting with one bare seed only.  This is certainly a
very interesting possibility, since we know that experiment has detected
many more light scalar mesons that can be accomodated in one nonet. 
We started by
investigating the $I=1$ sector, where two strong candidates have been
found: the long known $a_0(980)$ and the heavier $a_0(1450)$, detected by the
Crystal Barrel collaboration \cite{cbar}. By slightly increasing two crucial 
parameters of the model, the overall coupling $\gamma$ and the bare mass 
$m_0$, we have shown that it is possible to find a picture in which both 
states can easily be generated starting from one bare $n\overline n$ seed only.
Due to the structure of the vacuum polarization function, the heavier of the 
two states is automatically broader than the lighter one.
Drawing conclusions on further, heavier states would require detailed 
treatment of heavier thresholds, which are not included here.
 
\noindent
Encouraged by these results we then moved to the $I=1/2$ sector, to try and 
see whether we could also give a legitimate place to the controversial 
$\kappa$ meson. Due to the nature of the $n\overline s$ couplings to 
pseudoscalar-pseudoscalar channels, it turns out to be impossible to  
generate such a light $I=1/2$ scalar meson in our framework.
Further heavier states can be generated, one of which has the right features 
to be identified with the $K_0^*(1430)$, whereas the others might only be an 
artifact of the poor treatment of heavy thresholds in the model. Again, only 
a more rigorous description of such heavier thresholds could enable us to 
rule out their existence or not.

\noindent
To complete the study, we considered the $I=0$ sector: even though the 
heavily structured dynamics of the isoscalars make any clear--cut result 
far from robust, at least in the framework of our simple model, we can  
conclude that the multiple crossing mechanism is certainly active in this 
sector as well, and that dynamical generation of many states with the same 
quantum numbers but different masses can be a plausible explanation of the 
experimental occurence of many more $0^{++}$ states that can fit in one 
quark--model nonet.\\
We conclude that the detailed pole parameters of Ref.~\cite{roos} are 
strongly model dependent, as previously suggested by the comments of 
Refs.~\cite{isgur-speth}.

\noindent
As opposed to the work in Ref.\cite{sw}, we cannot tell which are intrinsic 
states and which are dynamically generated, nor can we state that we have 
definitely found enough strong candidates to complete two full nonets 
as in Refs. \cite{ochs,vanbev}: we can find no 
way to produce the light $\kappa$ meson and no precise assignment can be 
made for the isoscalar sector. As far as the isovector and 
isodoublet states are concerned, our picture is similar to that 
of Ref. \cite{ochs}, with the occurrence of two $a_0$ and two $K_0^*$ 
physical states and the possibility to produce a number of $f_0$'s, depending 
on the coupling scheme.

\section* {Acknowledgements}
This work was supported in part under the EU-TMR Programme, 
Contract No. CT98-0169, EuroDA$\Phi$NE.


\end{document}